1

# Void structure and cage dynamics in concentrated suspensions

## M. D. Haw[*]

School of Chemical, Environmental and Mining Engineering
University of Nottingham, University Park, Nottingham NG7 2RD, U.K.
and
School of Physics,
University of Edinburgh, Kings Buildings, Mayfield Road, Edinburgh EH9 3JZ, U.K.

---

[*] **Electronic address: mark.haw@nottingham.ac.uk**

**Abstract**

We analyse structure and dynamics in simulated high-concentration hard sphere colloidal suspensions by means of calculations based on the void space. We show that remoteness, a quantity measuring the scale of spaces, is useful in studying crystallization, since ordering of the particles involves a change in the way empty space is distributed. Calculation of remoteness also allows breakdown of the system into mesoscopic neighbor sets: statistics of mean remoteness and local volume fraction in these neighbor sets reveal that nuclei are formed at locally higher concentration, i.e. nucleation involves increased heterogeneity of the system. Full crystallization results in the transformation of the neighbor set mean remoteness distribution to an exponential form. The temporal fluctuation of local volume fractions in neighbor sets reveals significant details of dynamics, including abrupt dilations and compressions of local regions: leading to a clearer picture of the physical components of 'cage' dynamics in the colloidal glass.

# 1. Introduction

Examples of highly concentrated colloidal-scale particulate systems range from soils and building materials to foods, cosmetics and other advanced chemical products. The behaviour of these systems depends sensitively on their internal structure. Despite the fact that fundamental study of structure and dynamics in colloids has been a major theme of research over more than a century [1,2], substantial challenges remain, especially concerning highly concentrated suspensions. The development of model systems has played a significant role in advancing fundamental experimental study. For instance the debate over whether non-attractive 'hard sphere' colloids would form stable equilibrium crystals was settled experimentally by the observations of Pusey and van Megen [3]. Model polymeric hard sphere particles were indeed shown to crystallize at colloid volume fractions $\Phi > 0.49$, a purely entropic phase transition. Intriguingly, at higher volume fraction equilibrium crystallization was found to be suppressed, apparently associated with a dynamic 'glass transition' [4]: while the equilibrium state is crystalline, above $\Phi > 0.58$ crowding and drastic dynamic slowing prevents ordering on any reasonable experimental timescale. The structural and dynamic description of this colloidal glass transition has been the subject of exhaustive research over the past two decades. The most promising theoretical approach, so-called Mode Coupling Theory [5], remains difficult to interpret despite its many evident successes.

Characterising structure in concentrated, amorphous colloidal glasses, sediments and granular packings is itself a complex problem [6]. Traditional averaged methods such as pair correlation functions or their Fourier-space analogue (obtainable experimentally from scattering) are of limited use due to the lack of translational symmetry in amorphous colloidal glasses. Indeed, the dynamic colloidal glass transition seems not to be accompanied by any noticeable structural change as measured via light scattering experiments. Yet both the dynamics of glasses and the complete kinetic description of the crystallization transition, including its suppression at high $\Phi$, must depend on the details of local interparticle structure.

In this paper we demonstrate that examining the structure of the *empty space* in a concentrated particle system reveals useful information on structural transitions associated with crystallization, as well as providing intriguing details of local dynamics in a glassy suspension. We deliberately use very straightforward Monte Carlo methods to simulate diffusion in concentrated sphere-packings, as a simplest possible model system in which to demonstrate our structural analysis methods.

In what follows we first describe the basis of our structural analysis, give brief details of the Monte Carlo simulations, and then go on to discuss the results, concentrating on local structural effects in the hard sphere crystallization transition and on some initial investigations of local dynamics.

# 2. Methods

### A. Remoteness: analysis of local void structure

To measure local structure in a configuration of spherical particles, we start by considering the void or empty space not occupied by the particles. We define a function we term the *remoteness* of any point in the system, $R(\mathbf{r})$, which is equal to the distance from the point $\underline{r}$ to the nearest particle *surface*. For points inside particles, i.e. not part of the void, we define $R = 0$. As we shall see, the mean value of the function $R$ when measured over a whole system, as well as the distribution of $R$ and other local $R$-statistics, turn out to reveal significant structural and kinetic information.

Calculation of remoteness at a large number of randomly (uniformly) distributed points throughout a system naturally leads to the Voronoi construction, as shown in a 2D-schematic in Fig. 1(a): each void point is associated with the particle to whose surface it is closest, so that the set of void points closest to a given particle defines the Voronoi cell surrounding the particle. Calculation of remoteness is therefore an efficient alternative method of arriving at the Voronoi deconstruction of a system of particles. This provides an unambiguous definition of nearest neighbors (*i.e.* without the need for arbitrary distance cut-offs): nearest neighbors are those whose Voronoi cells share faces. It also allows us to define an aggregate volume of a 'neighbor set' associated with each particle, that is the sum of the particle's Voronoi cell plus those of its neighbors [Fig. 1(b)]. As we will show, this mesoscopic neighbor set volume provides a natural scale on which to examine local structural fluctuations, i.e. dynamics.

We note that what we call 'remoteness' has been used previously to study structure in fractal aggregation of colloids [7] and is also sometimes given the rather less expressive name '*P*' in statistical geometry texts [8]. The use of Voronoi methods, too, has a long history, especially in liquid structure [9] and granular matter [6], but also for instance in the study of protein structures [10], where the configuration and fluctuations of a protein are key to its function *e.g.* as an enzyme. Voronoi methods were used to identify neighbors in a recent molecular dynamics study of hard sphere crystallization [11], though the statistics of the void structure was not considered, the focus being rather on identifying crystal symmetries of nuclei.

### B. Monte Carlo simulations

In order to demonstrate the information that can be obtained from the void-space analysis we use a system consisting of a standard Monte Carlo simulation of diffusing equal-sized spheres. The particles occupy a cubic simulation box with periodic boundary conditions, at an overall system volume fraction Φ. The particles diffuse by random walk jumps selected from a uniform random distribution, with maximum jump set to 5% of the particle radius: a hard sphere interaction is imposed by rejecting any moves that cause overlap between the particles. Of course there exist far more advanced simulation techniques, and yet, for our purposes, this straightforward 'no-frills' Monte Carlo captures the essential features of diffusion in the concentrated colloidal system, allowing us to straightforwardly demonstrate the structural and dynamic features of the void space in the system.

The initial packing of spheres at given Φ is obtained using a 'swell and shuffle' algorithm: randomly distributed point (zero-radius) particles are diffused as their radii are increased, avoiding overlap, until the required starting Φ is obtained. We carry out runs of the simulation at a range of volume fractions Φ from 0.405 to 0.63. Simulations are run for an initial period of typically 50000-100000 attempted steps per particle in order to erase any possible structural transients due to the packing algorithm. After this period, measurements of structure evolution, dynamics etc, are commenced.

### 3. Results

#### A. Crystallization and mean remoteness

First we consider the broadest measure of void structure, the mean remoteness $<R>$ computed over the whole simulation system. $R$ is calculated at $10^6$ randomly selected points uniformly distributed across the system. In obtaining the mean $<R>$ we discard all points with $R=0$, *i.e.* all points inside particles, since the proportion of these points does not change with time (the overall volume fraction $\Phi$ is constant). We plot $<R>$ versus time (where in one timestep each particle makes one attempted move) for an example run at $\Phi=0.561$ [Fig. 2(a)]. We observe small fluctuations over an initial period, but then a significant drop in $<R>$, followed by another regime of small fluctuations around a constant value.

Example snapshots of this same system are shown in Figures 2(b)-(c), revealing that the steep drop in $<R>$ corresponds to *crystallization* of the particles. As mentioned, experiments in model hard sphere colloids also show entropy-driven ordering into single-phase hexagonal crystal at this volume fraction. But why should $<R>$ fall on crystallization? Though the *total volume* of empty space is of course conserved, remoteness measures the local *scale* of spaces. A consequence of crystallization is that the empty space becomes more evenly shared out than in the amorphous metastable system before crystallization (Figure 3). Amorphous structures tend to contain sets of neighbors closer to each other, and sets of neighbors further apart from each other, compared to the ordered system. Extra particles closer together results in a reduction of the population of small-$R$ void points, while extra particles further apart results in an increase in the population of large-$R$ void points. Fewer small-$R$ and more large-$R$ points means the amorphous system necessarily has a higher *mean remoteness* $<R>$.

Crystallization, in sharing out the void space more evenly, removes the largest-$R$ points and adds some small-$R$ points: hence the mean remoteness decreases. $<R>$ may thus be used as a kinetic measure of the extent of crystallization in a system, or indeed as a signal of the onset of crystallization.

#### B. Remoteness distribution: heterogeneity in crystallization

Next we consider the distribution of remoteness, $N(R)$, given by the proportion of points with remoteness $R$. Example distributions, once again for $\Phi=0.561$ for various times spanning the onset of crystallization, are shown in Figure 4. A noticeable population of larger $R$ points appears during crystallization, only to disappear again at later time once ordering is complete. Once again, measurement of remoteness can provide a 'signal' of ordering.

Why large-$R$ points, *i.e.* large voids, should appear during crystallization is not intuitively obvious. To demonstrate more clearly what is happening as the system crystallizes it is useful to consider the distribution of remoteness on an intermediate scale. In Figure 5 we plot the distribution of *neighbor set* mean remoteness $n(<R_{ns}>)$.

As mentioned above, we define the neighbor set volume associated with a given particle as the volume comprising its own Voronoi cell and the cells of its near neighbors, as shown in Fig. 1(b). $<R_{ns}>$ is then the mean remoteness for that neighbor set, calculated from all points within the given neighbor set volume (excluding points within the particles, as for $<R>$). $n(<R_{ns}>)$ is the distribution of neighbor set mean remoteness values across the whole system. This mesoscale distribution contains information on the heterogeneity of particles' local environments. It has the advantage that rather than involve an arbitrary 'coarse-graining' of the system into local zones, each neighbor set is naturally associated with the locality of a given particle and its non-arbitrarily defined neighbors.

The distribution $n(<R_{ns}>)$ reveals more clearly what is happening to the system as the particles order: there is a marked broadening of the distribution [*e.g.* compare Fig. 5(a), $t=100000$ with $t=900000$]. During crystallization the system becomes increasingly heterogeneous: some neighbor sets evolve to lower mean remoteness, others to higher, in other words some sets are compressing, some expanding. Intriguingly, as crystallization is completed an exponential tail develops in the $n(<R_{ns}>)$ distribution [Fig 5(a), $t=5\times10^6$]. This suggests that a wide range of variation across the neighbor set distribution characterises the ordered state: despite global order, the potential range of fluctuations, *i.e.* local disorder, is maximised. Assuming ergodicity (that the temporal distribution sampled by a single neighbor set is equal to the single-time ensemble distribution across the whole system as in Fig 5), once the particles have ordered, fluctuations of neighbor set mean remoteness can thus be described by a Boltzmann-like 'energetic' factor: the probability of fluctuations of a given set's $<R_{ns}>$ is given by the exponential distribution. Consistent with this, $n(<R_{ns}>)$ for a fluid system at $\Phi=0.405$ [Fig. 5(b)] also shows an exponential form on the RHS of the peak: the fluid is already at equilibrium.

### C. Local volume fractions

Definition of the neighbor set volume allows definition of a local volume fraction, $\Phi_{ns}$, once again in a non-arbitrary way associated with a given particle, based on each particle's local environment and set of neighbors. Figure 6(a) shows the distribution of $\Phi_{ns}$ across the system, $n(\Phi_{ns})$, for the simulation at overall $\Phi=0.561$ and times spanning the onset of crystallization. The broadening in $n(\Phi_{ns})$ during crystallization —the appearance of regions of lower and higher neighbor set volume fractions—is readily apparent. Once crystallization is complete these mesoscopic fluctuations in volume fraction disappear again, as expected at equilibrium where strong $\Phi$−gradients should not persist. Examination of particle configurations [Fig 6(b)-(c)] demonstrates that the high $\Phi_{ns}$ particles appearing during crystallization typically occur in regions of greater order: they represent compressed, ordered nuclei. Observations of crystallization kinetics by light scattering in model hard sphere colloids have been interpreted in a similar way: crystal nuclei initially form at densities significantly higher than the overall concentration, and 'decompress' as crystal growth proceeds [12].

Many simulation studies, in exploring the nucleation process [11, 13], have concentrated on identifying nucleation rates, the symmetry of nuclei (differentiating between fcc, bcc, icosahedral), etc. In principle a detailed comparison of remoteness distributions for these ideal crystalline forms might indeed reveal differences enabling discrimination of different symmetries: however this has not been pursued here. As often pointed out in the literature, whatever the 'ideal' forms of crystal order, in practice nuclei are rarely close enough to ideality to allow unambiguous identification. We are more interested, here, in considering the mesoscopic structural/kinetic signatures of the onset of order, rather than symmetries or quantitative calculation of rates of nucleation and growth.

### D. Limits to crystallization

In our simulations we observe crystallization at overall $\Phi$ as high as ~0.610, in contrast to experiments with model hard spheres where the glass transition intervenes at $\Phi$~0.58. Other simulations have also found crystallization at higher $\Phi$ than in experiments (indeed some simulations and experiments use polydisperse spheres deliberately in order to suppress crystallization). We do observe drastic slowing down of ordering as $\Phi$ increases, and by $\Phi$~0.610 we see no evidence of crystallization during > $10^7$ timesteps (an order of magnitude above the typical ordering time at $\Phi$=0.58). Of course crystallization may simply take so long that it becomes beyond the time resources available to us, but we note one intriguing point here concerning the behaviour of the distribution of neighbor set volume fractions during ordering. If broadening of $n(\Phi_{ns})$ is a necessary part of ordering [Fig 6(a)], then as $\Phi$ increases toward the random packing limit $\Phi_{RCP}$~0.64, the creation of regions at higher $\Phi_{ns}>\Phi$ must become increasingly difficult [Figure 6(d)]. (Once *ordered*, of course, such regions could in principle compress as high as $\Phi_{HCP}$=0.74, the maximum packing of hexagonally ordered monodisperse spheres—but they must order first!) *Whether nuclei can form in the densest regions may then be limited by the existence of the random packing limit*. The relation between the dynamic glass transition (which coincides *experimentally* with the observed suppression of order) and the random packing limit (true cessation of motion due to complete 'solid' packing) remains a matter of debate: the related question of how suppression of crystallization is related to the dynamic glass transition and to the existence of a random packing limit also deserves further study.

### E. Mesoscopic dynamics: local volume fraction fluctuations

There is a vast amount of data from dynamic scattering experiments and calculations on particle diffusion dynamics in colloidal glasses [2], yet such dynamic correlation functions remain intuitively difficult to interpret in terms of local structure. As an alternative, direct calculation of local concentrations *on a non-arbitrary scale* and associated with the locality of given particles, *i.e.* related to a non-arbitrary spatial structure, can reveal intriguing and intuitively appealing details of dynamics. The definition of neighbor sets and local neighbor set volume fractions allows us to study these mesoscopic density fluctuations in the system of spheres. In Figure 7 we plot the temporal fluctuations in neighbor set volume fraction for various overall $\Phi$, in each case for a single randomly chosen example particle. The trend with $\Phi$ is readily apparent. As $\Phi$ increases, fluctuations of $\Phi_{ns}$ slow down drastically (note the different time-scales in the plots of Figure 7). This is entirely unsurprising, but, more interestingly, at the higher $\Phi$ we observe the appearance of sudden jumps in $\Phi_{ns}$, overlaid on faster, smaller scale fluctuations. These jumps are essentially 'irreversible' shifts in local volume fraction: $\Phi_{ns}$ goes from fluctuating around a given long-lived 'plateau' to fluctuating around a new, also long-lived plateau [Fig 8(a), $\Phi$=0.610]. The jumps themselves become more obvious (plateaux more clearly separated) but also less frequent as $\Phi$ increases. At higher $\Phi$, local volume fractions are constrained to fluctuate on a smaller scale, with increasingly rare 'opportunities' for large fluctuations. In the limit of $\Phi_{RCP}$, local fluctuations become impossible.

Fig 8(b) shows $\Phi_{ns}$ for the same example particle as in Fig 8(a) at $\Phi=0.610$, plotted versus the particle's absolute displacement from an arbitrary starting point. The plot reveals a trajectory for the particle's neighbor set volume fraction and displacement that is broken into dense 'clumps': regions where $\Phi_{ns}$ and displacement fluctuate around a given value, interspersed with relatively sudden jumps between clumps. Dynamics in colloidal glasses is often discussed in terms of neighbor cages and cage-breaking events, where particles escape or otherwise change their local cage of neighbors. It would seem natural to associate the neighbor sets defined here with such cages, especially in view of the 'clumpy trajectory' in Fig 8(b). According to our results there is no simple definition of a cage-break: jumps between clumps in $\Phi_{ns}$-displacement space can involve separate diffusion and/or change in $\Phi_{ns}$, or indeed combinations of the two. The idea of a cage breaking event is not usually well defined in the literature: a more basic question is, how can we best describe the physical components of cage-dynamics? How do cages fluctuate? If cages and our neighbor sets are indeed taken as equivalent, Figure 8 shows that a description of cage-dynamics must include well-defined cage *dilations* and *compressions* as well as diffusion of the cage's constituent particles.

The observed neighbor set dilations/compressions may be associated with larger-scale fluctuations involving groups of neighbor sets: examining successively larger groups of neighbor sets would amount to investigation of dynamics on a range of scales yet still non-arbitrarily associated with given localities in the system. This approach may provide a more complete, quantitative and yet also clearly interpretable picture of the physical components of dynamics in concentrated particulate systems. Further work is in progress.

What is the origin of the long-lived 'plateaux' in $\Phi_{ns}$ between jumps, as seen in Fig 8(a)? What is it that allows a neighbor set suddenly to dilate (or indeed forces it to compress) and yet not to (immediately) reverse the change? One is tempted to propose a physical link between glassiness and 'jamming' at the mesoscale [14]: the jumps in $\Phi_{ns}$ represent dilative/compressive shifts between configurations that are then locally (temporarily) jammed against further large dilations or compressions. Faster, smaller-scale fluctuations (diffusion of the individual particles) eventually allow the local configuration to find another 'unjammed' state, at which point it dilates or compresses again—into another locally, temporarily jammed state. Interestingly, recent experiments and theory on the jamming of dense colloids under flow (or imposed stress) indicate the relevance of dilation/compression as part and parcel of jamming [15]. That dilations/compressions are also an important component of glassy dynamics, as suggested by our results, means that dilation/compression may be a useful concept with which to further illuminate the physical, microscopic nature of the link between glassy dynamics and 'jammed mechanics'.

## 4. Conclusion

We have demonstrated that examining local void structure in a model hard-sphere colloidal system at high concentration provides intriguing information on the crystallization process and on features of local structure and dynamics. Mean remoteness can be used as an indicator of the extent of ordering, while the definition of neighbor sets demonstrates how the local space in the crystallizing system evolves to a distribution allowing maximum local fluctuation. The neighbor set definition also allows 'deconstruction' of the system in a non-arbitrary way and calculation of mesoscopic concentrations associated, again non-arbitrarily, with the locality of given particles. Hence we have shown that, as overall volume fraction increases, the local mesoscale dynamics involves important dilations and compressions. Equating the neighbor set to the cage concept in glasses, we can begin to elaborate a better picture of cage-dynamics: mesoscopic dynamics involving combinations of diffusion, dilation and compression.

The study of void space and the breakdown of the system into neighbor sets has potential for advancing the description of dynamics and structure, in physically interpretable terms, in a range of dense particulate systems. For example, with recent advances in techniques such as confocal microscopy of colloids, these analyses could be carried out relatively straightforwardly on real colloidal suspensions [16]; while X-ray tomography and $\gamma$–ray absorption has recently been used to obtain direct structural data on granular packings [6], also amenable to remoteness analysis.

# Figure captions

**FIG. 1.** (a) Remoteness and the Voronoi construction. The shaded area around the central particle comprises all void points closest to that particle, i.e. its Voronoi cell. (b) Neighbor set definition. The neighbor set volume for the central particle comprises the volume of its Voronoi cell (unshaded) plus the volumes of the cells of its neighbors (shaded).

**FIG. 2.** (a) Mean remoteness $<R>$ *vs* time for a crystallizing system at $\Phi=0.561$. (b) Snapshot of the particle configuration at t=100000, i.e. before ordering. (c) Snapshot at $t=5\times 10^6$, after ordering.

**FIG. 3.** Ordering reduces mean remoteness. (a) Disordered system; (b) ordered system. The disordered system contains more large voids (points with large remoteness, arrowed in [a]). The ordered system contains fewer very close particles and so more small and medium-remoteness points (arrows in [b]).

**FIG. 4.** Remoteness distribution for $\Phi=0.561$, at times before (t=10000, 100000), during ($t=1.5\times 10^6$) and after ordering ($t=2\times 10^6$). Large-remoteness points appear during ordering.

**FIG. 5.** Mean remoteness in neighbor sets. (a) $\Phi=0.561$, for times before (t=100000), during (t=900000) and after ($t=5\times 10^6$) ordering. The dashed line is an exponential fit to the data for $t=5\times 10^6$ on the right-hand side of the peak. (b) $\Phi=0.405$. In the fluid, the distribution does not evolve in time, already exponential on the RHS of the peak (dashed line).

**FIG. 6.** Local volume fractions $\Phi_{ns}$ in neighbor sets. (a) Distribution for $\Phi=0.561$. During ordering (t=900000) the system becomes strongly inhomogeneous. (b) For the system in (a) at t=900000, snapshot of the particles at the centre of the most 'dilute' neighbor sets. (c) Particles in the most concentrated neighbor sets from (a), t=900000: these are also the most ordered. (d) $\Phi=0.610$, where no ordering is observed. For the distribution to spread, the most concentrated regions (where ordering would occur) would require $\Phi_{ns}>0.64$, i.e. above the random close packing limit.

**FIG. 7.** Temporal fluctuations of $\Phi_{ns}$ for a single example particle, for systems at various overall volume fractions $\Phi$. From top to bottom, $\Phi=0.405, 0.561, 0.585, 0.592, 0.601, 0.610, 0.630$.

**FIG. 8.** Fluctuations of $\Phi_{ns}$ at overall $\Phi=0.610$, for same particle as in Fig. 7, showing jumps in local volume fraction and in displacement of the central particle: the $\Phi_{ns}$-displacement trajectory (b) breaks into clumps.

**Figure 1.**

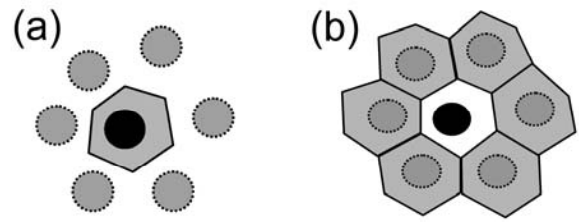

**Figure 2.**

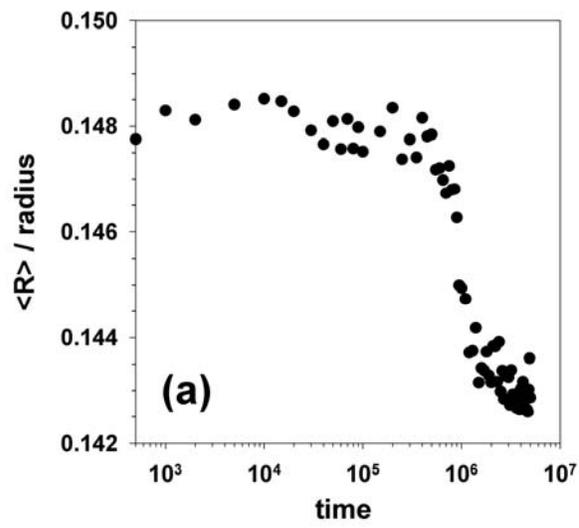

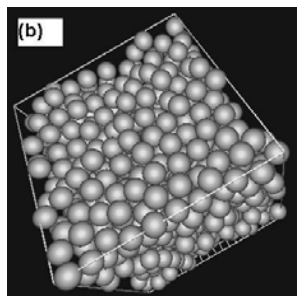
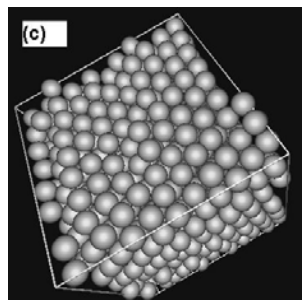

**Figure 3.**

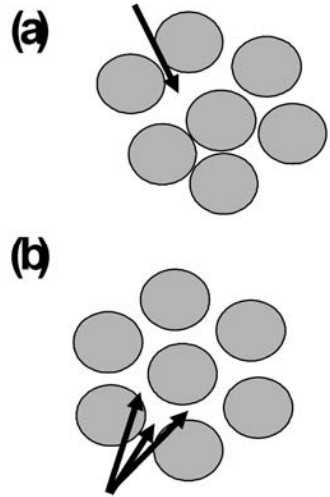

**Figure 4.**

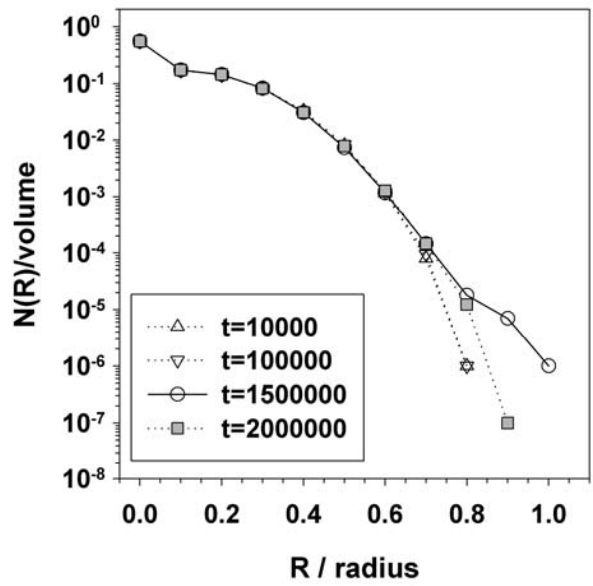

**Figure 5.**

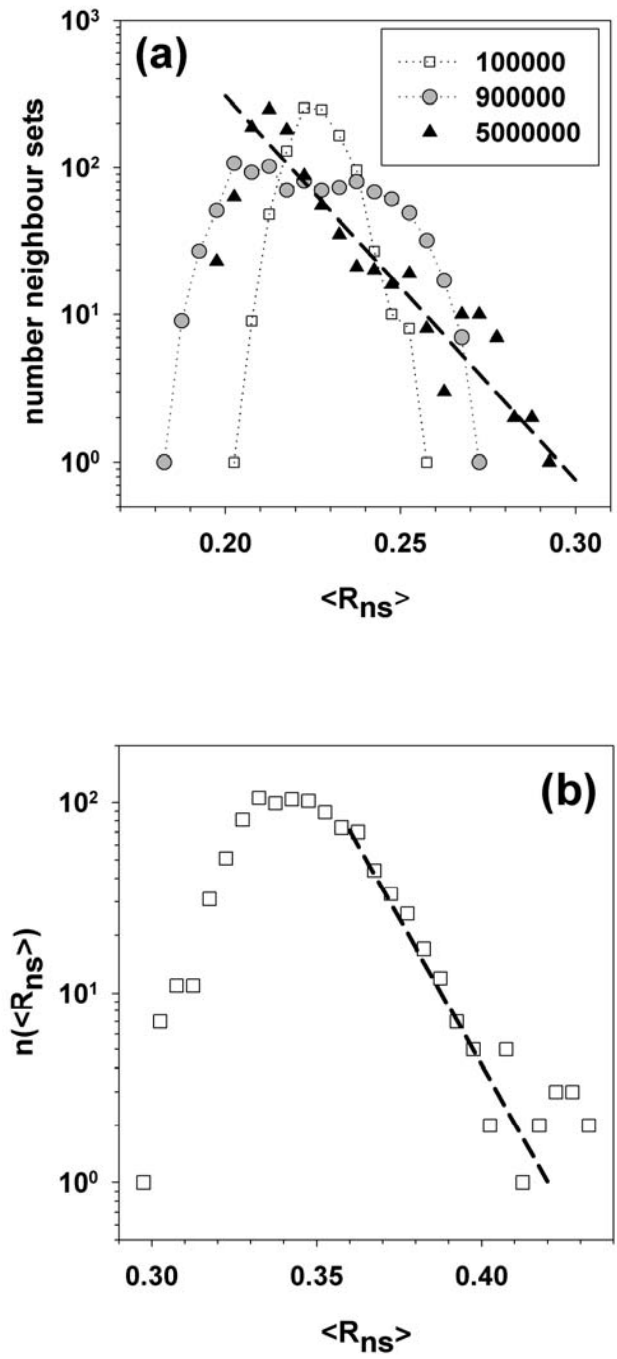

**Figure 6.**

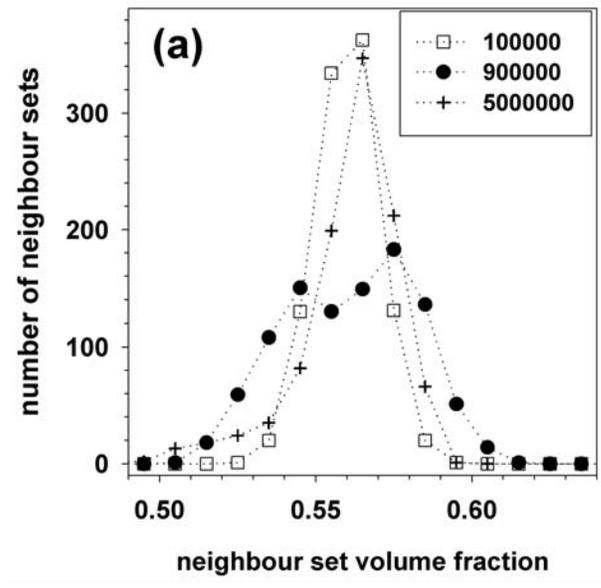

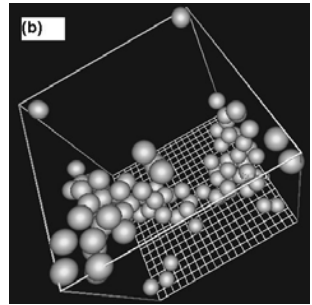 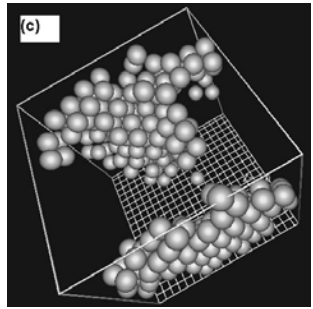

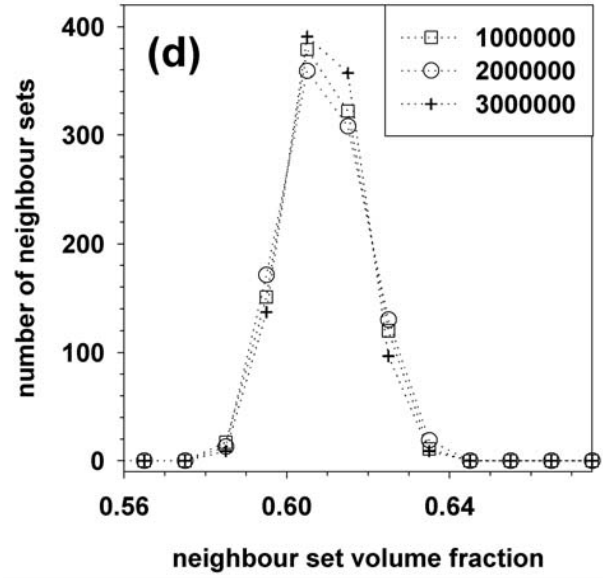

**Figure 7.**

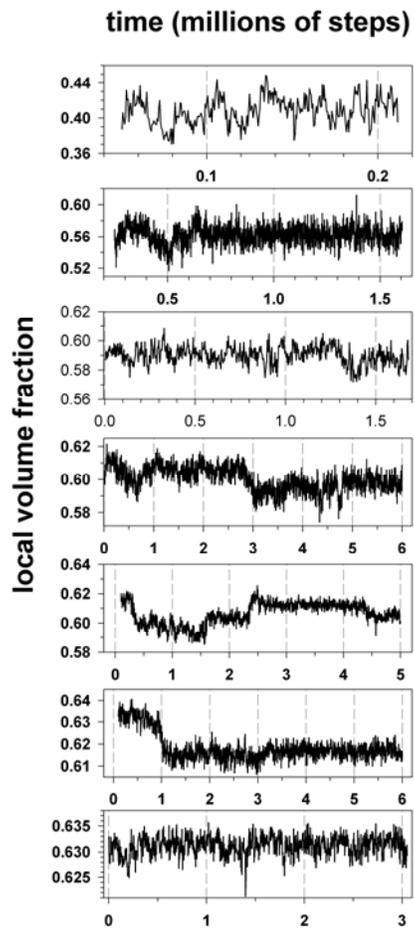

**Figure 8.**

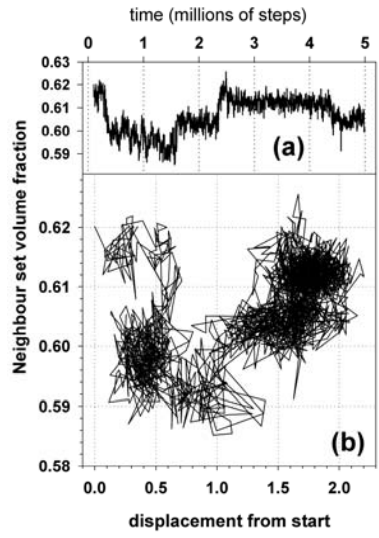